\documentstyle[twocolumn,aps]{revtex}
\newcommand{\noopsort}[1]{}

\begin{document}
\rm
\title{Quantum Computation in Brain Microtubules?\\ 
Decoherence and Biological Feasibility}
\author{S. Hagan \raisebox{3pt}{\scriptsize a} \and S. R. Hameroff 
\raisebox{3pt}{\scriptsize b,d} \and J. A. Tuszy\'nski 
\raisebox{3pt}{\scriptsize c,d}\\
\raisebox{3pt}{\scriptsize a} {\it Computational Modeling Lab, 
Department of Information Research, N.A.R.C.\\
3-1-1 Kannondai, Tsukuba, 305-8666, Japan}\\
\raisebox{3pt}{\scriptsize b} {\it Departments of Anesthesiology and 
Psychology and Center for Consciousness Studies\\
University of Arizona, Tucson, AZ 85724, USA}\\
\raisebox{3pt}{\scriptsize c} {\it Department of Physics, University 
of Alberta, Edmonton, Alberta, T6G 2J1, Canada}\\
\raisebox{3pt}{\scriptsize d} {\it Starlab NV/SA, Boulevard 
Saint-Michel 47, Brussels 1020, Belgium}}
\maketitle

\bigskip
\begin{abstract}
The Penrose-Hameroff (`Orch OR') model of quantum computation in 
brain microtubules has been criticized as regards the issue of 
environmental decoherence.  A recent report by Tegmark finds that 
microtubules can maintain quantum coherence for only $10^{-13}$ s, 
far too short to be neurophysiologically relevant.  Here, we 
critically examine the assumptions behind Tegmark's calculation and 
find that: 1) Tegmark's commentary is not aimed at an existing model 
in the literature but rather at a hybrid that replaces the superposed 
protein conformations of the `Orch OR' theory with a soliton in 
superposition along the microtubule, 2) Tegmark predicts decreasing 
decoherence times at lower temperature, in direct contradiction of 
the observed behavior of quantum states, 3) recalculation after 
correcting Tegmark's equation for differences between his model and 
the `Orch OR' model (superposition separation, charge vs. dipole, 
dielectric constant) lengthens the decoherence time to $10^{-5} - 
10^{-4}$ s and invalidates a critical assumption of Tegmark's 
derivation, 4) incoherent metabolic energy supplied to the collective 
dynamics ordering water in the vicinity of microtubules at a rate 
exceeding that of decoherence can counter decoherence effects (in the 
same way that lasers avoid decoherence at room temperature), and 5) 
phases of actin gelation may enhance the ordering of water around 
microtubule bundles, further increasing the decoherence-free zone by 
an order of magnitude and the decoherence time to $10^{-2} - 10^{-1}$ 
s.  These revisions bring microtubule decoherence into a regime in 
which quantum gravity can interact with neurophysiology.
\end{abstract}


\section{Introduction}

In the conventional biophysical approach to understanding cognitive 
processes, it has been generally accepted that the brain can be 
modeled, according to the principles of classical physics, as a 
neural network \cite{hop82,hop94,lit74,par86,kha90}.  Investigations 
in this field have delivered successful implementations of learning 
and memory along lines inspired by neural architectures and these 
have promoted optimism that a sufficiently complex artificial neural 
network would, at least in principle, incur no deficit in reproducing 
the full spectrum and extent of the relevant brain processes involved 
in human consciousness. 

However, physical effects in the functioning of the nervous system 
that lie outside the realm of classical physics suggest that such 
simulations may ultimately prove insufficient to the task.  One finds 
ample support for this in an analysis of the sensory organs, the 
operation of which is quantized at levels varying from the reception 
of individual photons by the retina \cite{lew88,bay79} to thousands 
of phonon quanta in the auditory system \cite{den89}.  Of further 
interest is the argument that synaptic signal transmission has a 
quantum character \cite{bec92,liu92}, although the debate on this 
issue has not been conclusive \cite{lar91}.  We note that using the 
thermal energy at room temperature in the position-momentum 
uncertainty relation, and assuming a {1 \AA} uncertainty for quantal 
effects, Beck and Eccles \cite{bec92} concluded that a particle whose 
mass is just six proton masses would cease to behave quantum 
mechanically and become classical for all intents and purposes.  This 
seems a serious underestimate, based on the de Broglie wavelength 
alone.  In any case, it is known that quantum modes of behavior exist 
in much larger structures such as peptides, DNA and proteins 
\cite{dav82}.  For instance, Roitberg {\it et al.} \cite{roi95} 
demonstrated functional protein vibrations that depend on quantum 
effects centered in two hydrophobic phenylalanine residues, and 
Tejada {\it et al.} \cite{tej96} have evidence to suggest that 
quantum coherent states exist in the protein ferritin.

The inadequacy of classical treatments is further suggested at the 
cognitive level, not only in regards to long-standing difficulties 
related to, for instance, accounts of semantics \cite{sea80}, binding 
\cite{jam90}, and the neural correlate of consciousness, but even in 
the rather modest goal of reproducing cognitive computational 
characteristics.  Penrose, in particular, has argued that human 
understanding must involve a non-computational element 
\cite{pen94,pen97} inaccessible to simulation on classical neural 
networks and this might be realized through a biological 
instantiation of quantum computation.\footnote{It has been noted that 
fundamentally analog mechanisms, based on continuous rather than 
traditional discrete (Turing) computation, might also constitute 
non-computation in the relevant sense and equally evade Penrose's 
argument.  It is not widely appreciated in the cognitive science 
community, however, that the essentially discrete nature of 
exocytosis implies that no such description can be framed in terms of 
the neurochemical basis of synaptic function.}  Along these lines, 
Penrose and Hameroff have put forth a specific 
model\cite{pen95,ham96b,ham98c} -- orchestrated objective reduction 
(Orch OR) -- positing quantum computation in microtubule protein 
assemblies in the neurons of the brain. 

Microtubules are hollow cylinders whose walls consist in 13 columns 
(protofilaments) of the protein tubulin arranged in a skewed 
hexagonal lattice.  Along with other structures, microtubules 
comprise the internal scaffolding -- the `cytoskeleton' -- in cells 
including neurons.  Determinants of both structure and function, 
cytoskeletal structures are dynamically active, performing a host of 
activities instrumental to cellular organization and 
intelligence\cite{dus84}.  Earlier models (see, for instance, 
\cite{ham82,ham87,ras90,sat93,day94,tus95,tus97b}) proposed classical 
information processing among the tubulin `dimers' composing 
microtubules -- molecular-level automata regulating real-time 
cellular behavior.  More recently, arguments have been made for 
quantum computation at the level of individual proteins 
\cite{con94}.  In particular, functional protein conformational 
states are mediated by quantum van der Waals forces\cite{voe95}, the 
relevance of which is demonstrated by the mechanism of action of the 
general anesthetic gases that reversibly ablate consciousness.  
Anesthetics act by disturbing such forces in the hydrophobic pockets 
of various brain proteins \cite{fra84,ham98b}.  Microtubules are thus 
poised to mediate between a tubulin-based quantum computation and the 
classical functioning of neurons.

The Penrose-Hameroff proposal suggests that coherent superpositions 
of tubulin proteins are inherently unstable and subject to 
self-collapse under a quantum gravitational criterion.  Calculations 
indicate that decoherence due to such a quantum gravitational 
mechanism might allow coherence to survive for milliseconds, even up 
to a second, in the context of the brain 
\cite{pen95,ham96b,mav97,mav98}.  According to the Orch OR model, 
quantum coherent states that persist over such a neurophysiologically 
relevant time frame could influence cognitive processes, in a manner 
accounting for the non-computational element, by `orchestrating' 
state vector reductions to perform quantum computation.  

Recently, Tegmark \cite{teg00} has responded to this and other models 
of brain function invoking a quantum element by contending that the 
relevant degrees of freedom cannot reasonably be sufficiently 
shielded from environmental, and particularly thermal, influence to 
maintain quantum coherence until self-collapse.  It is well-known 
that technological quantum devices often require extremely low 
temperatures to avoid decoherence through environmental interaction.  
The survival of a delicate quantum coherence in the `warm, wet and 
noisy' milieu of the brain long enough for quantum computation to 
play a neurophysiological role therefore seems unlikely to many 
observers.  Tegmark maintains that orthodox mechanisms of decoherence 
would collapse a microtubule-associated quantum state on a timescale 
of the order of $10^{-13}$ seconds, much shorter than that associated 
with decoherence by quantum gravitational means, thus apparently 
superseding the possibility of an orchestrated reduction relevant to 
neurophysiology.\footnote{Tegmark also maintains that the original 
motivations for the modifications to the Schr\"odinger equation 
implicit in a reduction postulate are obviated by the Everett 
relative state interpretation of quantum theory \cite{eve57,eve73}.  
As has been previously noted \cite{des76,boh93}, however, Everett's 
determination that the probability interpretation of the diagonal 
elements in the density matrix automatically holds for most observers 
depends on the prior assumption of a probability interpretation for 
the measure over the space of observers, and so itself constitutes an 
additional postulate.}

In the following, we critically review Tegmark's assumptions, 
calculations, and claims to ascertain whether they accurately reflect 
the potential for quantum computation in the brain. 

\section{Decoherence Rates}

Tegmark considers in his paper \cite{teg00} two different scales at 
which quantum computation might occur in the brain -- one involving 
superpositions of neurons firing and not firing (with calculated 
decoherence times of $10^{-20}$ seconds), and another involving 
microtubules (calculated decoherence times of $10^{-13}$ seconds).  
We agree that superpositions at the level of neural firing are 
unlikely, and in fact play no role in the Orch OR or any other 
contemporary quantum model.  We therefore focus our attention on 
Tegmark's assertions regarding decoherence times for 
microtubule-associated quantum states.  

Though Tegmark specifically implicates Penrose, his criticisms target 
neither the Penrose-Hameroff Orch OR model nor any other that is 
currently or has been under investigation.  It appears to be directed 
against a spuriously quantum version of a classical model, put forth 
by Sataric {\it et al.} \cite{sat93}, to treat lossless energy 
transfer in microtubules in terms of kink solitons travelling along 
their length.  

Tegmark considers a model in which kink soliton solutions, like those 
of Sataric {\it et al.} \cite{sat93}, exist in a quantum 
superposition of different positions along the microtubule.  The 
actual Orch OR model, on the other hand, is framed in terms of 
superpositions of the conformational state of a tubulin dimer.  The 
fact that this state is coupled to a delocalized electron residing in 
the hydrophobic pocket of the tubulin dimer protein points to a 
process of conformational change in the dimer controlled by quantum 
level events.  There is thus a considerable conceptual disparity 
between this model and that considered by Tegmark.  Nevertheless, it 
is equally critical to the actual model that the mechanisms of 
decoherence analyzed do not destroy quantum coherence before a 
quantum gravity-induced self-collapse can come into play.  Below, we 
consider both numerical and theoretical concerns that bear on the 
results presented by Tegmark.  

In the microtubule case, Tegmark determines the time to decoherence, 
$\tau$, due to the long-range electromagnetic influence of an 
environmental ion to be:
\begin{equation}
\tau\sim\frac{4\pi\epsilon_0 a^3\sqrt{mkT}}{Nq_e^2 s},
\label{eq:decohere}\end{equation}
where $T$ is the temperature, $m$ is the mass of the ionic species, 
$a$ is the distance to the ion from the position of the quantum 
state, $N$ is the number of elementary charges comprising that state, 
and $s$ is the maximal `separation' between the positions of the 
tubulin mass in the alternative geometries of the quantum 
superposition.  Since any difference in the mass distributions of 
superposed matter states will impact upon the underlying spacetime 
geometry, such alternative geometries must presumably be permitted to 
occur within the superposition.

It is worthy of immediate note that the decoherence time given by 
Tegmark in equation (\ref{eq:decohere}) is manifestly incorrect with 
respect to its temperature dependence.  As absolute temperature 
increases, the opposite should happen to the decoherence time as a 
result of the increasing influence of environmental fluctuations on 
the quantum rate.  This will be further elaborated in section 
\ref{sec:lowtemp}, but we first consider those  corrections required 
by the current formulation that do not alter the theoretical 
foundation of the derivation.

\subsection{Superposition `separation'}

Superposition occurs not only at the level of a mass distribution 
separated from itself, but concomitantly at the level of the 
underlying spacetime geometry.  According to Penrose's quantum 
gravitational criterion for objective reduction, superpositions 
involving different spacetime geometries are considered inherently 
unstable, with the rate of collapse determined by a measure of 
difference in the geometries.  As this measure approaches the order 
of a Planck length, it becomes problematic to determine a consistent 
standard by which to match up points in the superposed geometries.  
Yet if the superposed spaces cannot be resolved into one and the same 
space, then the different matter states in the superposition must 
occur in separate spaces and the meaning of `superposition' in this 
context becomes obscure.  Thus the rate of collapse in Penrose's 
suggestion for objective reduction must become significant before the 
measure of difference in superposed spacetime geometries grows to the 
Planck scale.  Since gravitational forces are inherently weak, 
however, the mass distributions of the superposed matter states can 
be substantially `separated' before incurring a large measure of 
difference in the associated spacetime geometries.  

Tegmark assumes that this separation, $s$, must be at least as large 
as the diameter of a microtubule, $D = 24$ nm, for superpositions 
spanning many tubulin dimers.  This estimate is based on a picture of 
tubulin dimers literally `beside themselves' in superposition.  
However, in the Orch OR theory, the authors contemplate separation 
only at the level of the individual atomic nuclei of amino acids 
comprising the protein.  

Assuming that a conformational movement of tubulin displaces its mass 
by roughly one tenth the radius of a tubulin monomer, Hameroff and 
Penrose \cite{ham96a} surveyed three different levels at which 
separation might occur: 1) partial separation (10\%) of protein 
spheres, 2) complete separation of atomic nuclei, and 3) complete 
separation of nucleons.  The gravitational self-energy in each 
instance is taken to be inversely proportional to the decoherence 
time according to the energy-time uncertainty relation.  In the case 
of protein spheres, the energy, $E$, for partial separation is 
obtained from:
\begin{equation}
E=\frac{GM^2s^2}{2r^3} \left( 
1-\frac{3s}{8r}+\frac{s^3}{80r^3}\right) ,
\nonumber
\end{equation} 
where $M$ is the monomer mass of 55 kDa, $r$ is the radius of a 
monomer sphere, and $s=\frac{\scriptsize 1}{\scriptsize 10}r$ is the 
superposition separation.  For complete separations at the level of 
either atomic nuclei or nucleons, the contribution to the self-energy 
determined in separating the mass distributions to a distance of one 
diameter (the contact position in a spherical approximation of the 
masses) is of the same order as that determined by increasing the 
separation further, even to infinity, so the contribution in moving 
from coincidence to contact is a good order of magnitude estimator of 
the self-energy for complete separations.  

Mass separation of granular arrays of atomic nuclei yields the 
highest energies of the three cases, and hence the shortest 
decoherence times, and it is this level that will thus dominate in an 
orchestrated reduction.\footnote{Estimates of the time to decoherence 
due to such a quantum gravitational mechanism will depend on the 
number of tubulin subunits participating in the quantum state.  For 
example, calculating energies based on a separation at the level of 
atomic nuclei, a decoherence time of 500 ms is obtained for $10^9$ 
participating tubulin, or about $10^3$ neurons if it is assumed that 
$10\%$ of the tubulin contained becomes coherent.  Larger quantum 
states will collapse more rapidly.}  Thus mass separation is effected 
already at separations the size of the nucleus, on the order of 
femtometers, some seven orders of magnitude smaller than Tegmark's 
estimate.  

\subsection{Polarization and charge}

In his analysis of the polarization associated with the microtubule, 
Tegmark defines $p(x)$ to be the average component, in the direction 
parallel to the microtubule axis, of the electric dipole moment due 
to the tubulin dimers, a polarization function given in units of {\it 
charge} $\times$ {\it length}.  He then claims that $-p^\prime (x)$ 
represents the net charge per unit length along the microtubule, 
which, on dimensional grounds alone, cannot be well-founded.  
Nevertheless, on this basis, he integrates over the length of the 
microtubule across the kink to obtain a net charge that incorrectly 
bears the units of an electric dipole moment.  This, in effect, 
treats the microtubule as a line of uniform charge rather than a 
polarized line, and this is how he obtains the magnitude of the {\it 
polarization} function by simply summing the {\it charge} of the ions 
arrayed around the microtubule at the level of the kink-like 
propagation.  The value of $N$ that figures in his estimate of the 
decoherence time is then this sum expressed in units of the electron 
charge, $q_e$.  Aside from the dimensional incongruities in this 
procedure, Tegmark accounts only for the presence of 18 ${\mathrm 
Ca}^{2+}$ ions, bound to the C-terminus of the tubulin on each of 13 
protofilaments in a cross-section of the microtubule.  This overlooks 
the negative charges borne by amino acid side groups and numerous 
other charges associated with tubulin, all of which attract 
counterions from the surrounding medium.  

Tubulin has only been imaged to atomic resolution within the last two 
years, following twenty years of difficult work with this protein.  
Nogales {\it et al.} published the structure of $\alpha$- and 
$\beta$-tubulin, co-crystallized in the heterodimeric 
form\cite{nog98}.  The work establishes that the structures of 
$\alpha$- and $\beta$-tubulin are nearly identical and confirms the 
consensus speculation.  A detailed examination shows that each 
monomer is formed by a core of two $\beta$-sheets that are surrounded 
by $\alpha$-helices.  The monomer structure is very compact, but can 
be divided into three functional domains: the amino-terminal domain 
containing the nucleotide-binding region, an intermediate domain 
containing the taxol-binding site, and the carboxy-terminal 
(C-terminus) domain, which probably constitutes the binding surface 
for motor proteins\cite{nog98}.

Recently, tubulin's electrostatic properties, including its potential 
energy surface, were calculated \cite{bro90} with the aid of the 
molecular dynamics package {\sc tinker}.  This computer program 
serves as a platform for molecular dynamics simulations and includes 
a facility to use protein-specific force fields.  With the C-terminus 
tail excluded, the electrostatic properties of tubulin are summarized 
below, following Brown\cite{bro90}.

\begin{center}
\begin{tabular}{|c|c|c|}
\hline
\multicolumn{2}{|c|}{Tubulin Property}&$\quad$ Calculated Value 
$\quad$\\
\hline\hline
\multicolumn{2}{|c|}{Charge}& -10 $q_e$\\
\hline
\multicolumn{2}{|c|}{Dipole Moment}& 1714 Debye\\
\hline
Dipole&$\quad p_x\quad$& 337 Debye\\
\cline{2-3}
Moment&$\quad p_y\quad$& -1669 Debye\\
\cline{2-3}
$\quad$ Components $\quad$&$\quad p_z\quad$& 198 Debye\\
\hline
\end{tabular}
\end{center}

\vspace{-0.05in}

{\footnotesize{\bf TABLE I.} Calculated values of some electrostatic 
properties of tubulin.}

\vspace{0.1in}

Since $1 {\mathrm\; Debye}=\frac{\mbox{\rm\scriptsize 
1}}{\mbox{\rm\scriptsize 3}}\times 10^{-29}\,{\mathrm C}\cdot 
{\mathrm m}$, we find that the total dipole moment is approximately 
$5.7\times 10^{-27}\,{\mathrm C}\cdot {\mathrm m}$, but only a fifth 
of it is oriented along the protofilament axis.\footnote{The 
$x$-direction coincides with the protofilament axis.  The $\alpha$ 
monomer is in the direction of increasing $x$ values, relative to the 
$\beta$ monomer.  This is opposite to the usual identification of the 
$\beta$ monomer as the `plus' end of the microtubule, but all this 
identifies is whether the microtubule is pointed towards or away from 
the cell body.}

It turns out that tubulin is quite highly negatively charged at 
physiological pH, but that much of the charge is concentrated on the 
C-terminus.  This is the one portion of the tubulin dimer which was 
not imaged by Nogales {\it et al.} \cite{nog98} due to its freedom to 
move following formation of the tubulin sheet.  This tail of the 
molecule extends outward away from the microtubule and into the 
cytoplasm and has been described by Sackett \cite{sac95}.  At neutral 
pH, the negative charge on the carboxy-terminus causes it to remain 
extended due to the electrostatic repulsion within the tail.  Under 
more acidic conditions, the negative charge of the carboxy-terminal 
region is reduced by associated hydrogen ions.  The effect is to 
allow the tail to acquire a more compact form by folding.

Any exposed charge in a cytoplasm will be screened by counterions 
forming a double layer.  The screening distance provided by these 
counterions and water is the Debye length and, in the case of 
microtubules, its value is typically 0.6-1.0 nm under physiological 
conditions.  Due to the exposure of negatively charged amino acids in 
the C-terminus, a Debye layer is formed, screening thermal 
fluctuations due to the stronger Coulomb interactions over distances 
within the Debye length.

Ionic forces thus tend to cancel over even relatively short distances 
so that the forces mediating between tubulin and its environment 
should instead be characterized by dipolar interactions.  This 
suggests that Tegmark's derivation of the decoherence time in 
equation (\ref{eq:decohere}) should be replaced with one that 
characterizes tubulin in terms of its electric dipole moment, thus 
avoiding the need to make a rather arbitrary cut in selecting which 
charges are to be constitutive of the overall charge of the kinked 
microtubule and which are to be neglected.  Such a modification is 
accomplished by replacing the Coulomb potential, 
$V_{\mbox{\rm\footnotesize Coulomb}}=q^2/4\pi\epsilon_0\vert {\mathbf 
r}_1-{\mathbf r}_0\vert$, describing the interaction of a quantum 
state of charge $q$ at ${\mathbf r}_0$ and a similarly charged 
environmental ion at ${\mathbf r}_1$, in favor of a dipole potential, 
$V_{\mbox{\rm\footnotesize dipole}}=q{\mathbf p}\cdot ({\mathbf 
r}_1-{\mathbf r}_0)/4\pi\epsilon_0\vert {\mathbf r}_1-{\mathbf 
r}_0\vert^3$, parameterized by ${\mathbf p}$, the electric dipole 
moment due to tubulin of the kinked microtubule.  The interaction is 
well-approximated, for the purposes of an order-of-magnitude 
estimate, by this dipole potential in the case that $a$ is greater 
than the separation of charges in the determination of the electric 
dipole moment.  This separation will not generally be larger than the 
length of a tubulin dimer, 8 nm, whereas $a=\frac{\mbox{\scriptsize 
1}}{\mbox{\scriptsize 2}}D+n^{-1/3}\approx 14$ nm for the same ionic 
density used by Tegmark, $n=\eta n_{H_2O}$ with $\eta\approx 2\times 
10^{-4}$. 

As in the Coulomb case of interacting charges, the force resulting 
from the dipole potential contributes only a phase factor in the 
evolution of the (reduced) density matrix, traced over the 
environmental degrees of freedom.  It is thus tidal effects that 
determine the leading contribution to the rate of decoherence.  In 
terms of the vector ${\mathbf a}\equiv {\mathbf r}_1^0 -{\mathbf 
r}_0^0$, between the initial average positions of the environmental 
ion and the polarized quantum state, these tidal effects are given by 
the Hessian matrix of second derivatives of the interaction potential:
\begin{equation}
{\mathbf M}=\frac{3 q p}{4\pi\epsilon_0 a^4}\left[ (5\hat{\mathbf 
a}\hat{\mathbf a}^T-{\mathbf I})(\hat{\mathbf p}\cdot\hat{\mathbf 
a})-  (\hat{\mathbf a}\hat{\mathbf p}^T + \hat{\mathbf p}\hat{\mathbf 
a}^T)\right] .
\label{eq:hessian}
\end{equation}

Under the same assumptions that give rise to equation 
(\ref{eq:decohere}), the dipole case yields a decoherence timescale of
\begin{equation}
\tau\sim\frac{4\pi\epsilon_0 a^4\sqrt{m k T}}{3 q_e p 
s}\Omega_{\mbox{\rm\footnotesize dipole}},
\label{eq:dipole}
\end{equation}
where 
\begin{eqnarray}
\Omega_{\mbox{\rm\footnotesize dipole}}&=& \Bigl( 
5\cos^2\theta\cos^2\varphi -4\cos\theta\cos\varphi\cos\psi
\nonumber\\ 
&+&\cos^2\theta +\cos^2\varphi+\cos^2\psi  \Bigr)^{-\frac{1}{2}},
\nonumber
\end{eqnarray}
is a geometric factor fixed in terms of the angles between ${\mathbf 
p}$, ${\mathbf s}$ and ${\mathbf a}$: 
\begin{eqnarray}
\cos\theta&=&\hat{\mathbf a}\cdot\hat{\mathbf s},\nonumber\\
\cos\varphi&=&\hat{\mathbf p}\cdot\hat{\mathbf a},\nonumber\\
\cos\psi&=&\hat{\mathbf s}\cdot\hat{\mathbf p}.
\nonumber
\end{eqnarray}
In our calculations, $\Omega_{\mbox{\rm\footnotesize dipole}}$ is 
taken to be of order one.\footnote{Though 
$\Omega_{\mbox{\rm\scriptsize dipole}}$ increases without bound as 
the three vectors, $\hat{\mathbf a}$, $\hat{\mathbf p}$, and 
$\hat{\mathbf s}$,  approach mutual orthogonality, randomly oriented 
vectors rarely come close enough to satisfying this condition to make 
an order of magnitude difference in the decoherence time.}

The calculation of the decoherence timescale in equation 
\ref{eq:dipole} can be made more realistic by taking into account the 
dielectric permittivity of tubulin in cytosol, neglected in the 
original calculation.  Since the intracellular medium is primarily 
water, its dielectric constant can be quite high.  The precise value 
of the permittivity of water is both temperature and frequency 
dependent but can be as high as $\epsilon\approx 
80$\cite{has73,tus97c}.  Conservatively estimating the dielectric 
constant of the surrounding medium by $\epsilon \approx 10$, and 
using the values, determined above, for the component of tubulin's 
electric dipole moment along the microtubule axis yields a 
decoherence time, $\tau\approx 10^{-5}-10^{-4}$ s, that is already 
eight or nine orders of magnitude longer than that suggested by 
Tegmark.

We also wish to point out that Mavromatos and Nanopoulos \cite{mav98} 
estimated decoherence times for dipolar excitations in microtubules.  
Depending on the set of assumptions adopted, the value of $\tau$ 
obtained ranged from as low as $10^{-10}$ s using a conformal field 
theory method to as high as $10^{-4}$ s using a coherent dipole 
quantum state.  For a kink state similar to that discussed by 
Tegmark, that value is on the order of $10^{-7}-10^{-6}$ s.

\subsection{Dynamical timescales, shielding and error correction}

Given the sizeable discrepancy between these estimates and those of 
Tegmark, it seems reasonable to re-evaluate whether the assumptions 
made in his calculation of decoherence rates remain valid.  In 
particular, the derivation requires that the decoherence timescale 
should fall far short of any relevant dynamical timescale for either 
the quantum object or the ionic environment, if the non-interacting 
contribution to the Hamiltonian is to be neglected relative to the 
interaction contribution.  With the substantially modified 
decoherence times calculated above, this assumption is no longer 
justified, even by Tegmark's own estimates which place the dynamical 
timescale for a kink-like excitation traversing a microtubule at 
$\tau_{\mbox{\footnotesize\rm dyn}}\approx 5\times 10^{-7}$ s.  
Indeed, the requirement may not even be met over the dynamical 
timescale of neurons firing, a scale that Tegmark places in the range 
$\tau_{\mbox{\footnotesize\rm dyn}}\approx 10^{-4} - 10^{-3}$ s.

Two possible avenues might be envisioned in the framework of the Orch 
OR theory by which to overcome the influence of decoherence due to 
scattering and tidal effects, such that decoherence by quantum 
gravitational effects might play a role.  The most obvious solution 
is to require that the shortest decoherence times be those due to 
quantum gravity. An equally viable approach, however, is to require 
that decoherence due to other mechanisms be effectively countered by 
dynamical processes operating on timescales more rapid than that of 
the relevant form of decoherence.  This is the means by which quantum 
systems like lasers maintain quantum coherence against thermal 
disruption at biologically relevant temperatures.  The dynamical 
timescale is here determined by the rate at which the system is 
pumped by an incoherent source of energy.  Appropriate dynamical 
timescales in the microtubule case might be determined, for instance, 
by the characteristic rate at which the incoherent energy provided by 
GTP hydrolysis -- known to control the stability of microtubules 
\cite{hym92,hym95} -- is supplied to processes that maintain the 
quantum state against decoherence by scattering, such as actin 
gelation in sol-gel cycles and the ordering of water.  

The transition between the alternating phases of solution and 
gelation in cytoplasm depends on the polymerization of actin, and the 
particular character of the actin gel in turn depends on actin 
cross-linking.  Of the various cross-linker related types of gels, 
some are viscoelastic, but others (e.g. those induced by the actin 
cross-linker avidin) can be deformed by an applied force without 
response \cite{wac94}.  Cycles of actin gelation can be rapid, and in 
neurons, have been shown to correlate with the release of 
neurotransmitter vesicles from pre-synaptic axon 
terminals\cite{miy95,mua95}. In dendritic spines, whose synaptic 
efficacy mediates learning, rapid actin gelation and motility mediate 
synaptic function, and are sensitive to anesthetics 
\cite{kae97,fis98,kae99}.

In the Orch OR model, actin gelation encases microtubules during 
their quantum computation phase.  Afterwards, the gel liquifies to an 
aqueous form suitable for communication with the external 
environment.  Such alternating phases can explain how input from and 
output to the environment can occur without disturbing quantum 
isolation.

The water within cells is itself not truly liquid, but has been shown 
to be, to a large extent, ordered\cite{cle84}.  Most of the ordered 
water in the cell in fact surrounds the cytoskeleton \cite{cle81}.  
Neutron diffraction studies indicate several layers of ordered water 
on such surfaces, with several additional layers of partially ordered 
water.  Tegmark himself allows that the dynamical process of ordering 
water in the vicinity of the microtubule\footnote{Curiously, while 
the point is conceded with respect to the water {\it inside} the 
microtubule, Tegmark finds it more contentious as regards the water 
{\it outside} the microtubule, which ``fills the entire cell 
volume.''  The mechanism of ordering is independent of whether the 
water is inside or outside the microtubule, and is only contended for 
the water closely approaching the microtubule.} could protect the 
quantum system from short-range sources of decoherence such as the 
scattering of nearby molecules.  

In fact, there is a long history to the hypothesis that macroscopic 
quantum coherence might be supported biologically by maintaining a 
supply of energy at a rate exceeding a threshold value 
\cite{fro68,fro70,fro75}.  The collective effects responsible for the 
ordering of water arise in the context of a supply of metabolic 
energy \cite{cle84}.  Empirical evidence indicates that, in the 
presence of an activation energy approximating the amount required 
for the formation of a soliton on the microtubule ($\approx 0.3$ eV), 
the surrounding water can be easily brought to an electret state 
\cite{cel77,has81}.  Spontaneous breaking of the dipole rotational 
symmetry in the interaction of the electric dipole moment of water 
molecules with the quantized electromagnetic field would then give 
rise to the dipolar wave quanta that are postulated to mediate 
collective effects \cite{del85,del86,jib94}. 

In the gel phase, the water-ordering surfaces of a microtubule
are within a few nanometers of actin surfaces which also order water.
Thus bundles of microtubules encased in actin gel may be effectively 
isolated with the decoherence-free zone, $a$, extending over the 
radius of the bundle, on the order of hundreds of nanometers.  
Applied to
the previously corrected version of Tegmark's equation, an order of 
magnitude increase in the the decoherence-free zone results in a 
revised decoherence time for the microtubule bundle of $10^{-2} - 
10^{-1}$ s.

Technological quantum computing is, in general, feasible because of 
the use of quantum error correction codes.  It has been suggested 
that error correction may be facilitated by topologies -- for 
instance, toroidal surfaces \cite{kit97,bra98}-- that allow 
computation and error correction to run along different axes, 
repeatedly intersecting.  Similarly, quantum computation in the 
medium of microtubules may proceed longitudinally along 
protofilaments, with error correction codes running around the 
microtubule circumference in helical pathways.  Interpenetration of 
the left- and right-handed pathways occurs such that the numbers of 
rows in the two pathways are successive Fibonacci numbers.  
Penrose\cite{pen99} has suggested that this might be optimal for 
quantum error correction.

\subsection{Low temperature limit}
\label{sec:lowtemp}
  
An examination of limiting cases casts further doubt on the validity 
of the reasoning that led Tegmark to claim such a rapid decoherence 
rate due to long-range forces.  If the adoption of equation 
(\ref{eq:decohere}), even in the modified form, (\ref{eq:dipole}), is 
justified and definitively forecasts the climate for coherence, then 
it would appear that no quantum coherent states are likely to exist 
at {\it any} temperature.  Both equations require that, as the 
temperature approaches absolute zero, decoherence times should tend 
to zero as the square root of temperature.  The apparent implication 
is that low temperatures, at which decohering environmental 
interactions should presumably have minimal impact, are deemed most 
inhospitable to the formation and preservation of quantum coherence, 
contrary to experience.  

Yet the low temperature regime is precisely the context in which the 
assumptions on which Tegmark premises his argument should be most 
clearly valid.  Both object and environment should be well-localized 
near their initial average positions in this limit and, unless it is 
imagined that the dynamical timescale goes to zero in the low 
temperature limit even more rapidly than the decoherence scale -- 
entailing a dynamical rate that increases without bound as absolute 
zero is approached -- the requirement that the decoherence scale lie 
well below the dynamical timescale is also met.  Accounting for the 
temperature dependence implicit in $a$, which must decrease to a 
minimum in the absence of thermal agitation, only exacerbates the 
counter-intuitive trend. 

As quantum coherent states do, in fact, exist, and the predictions of 
equation (\ref{eq:decohere}) run contrary to observation, Tegmark's 
conclusions appear unfounded.  

\section{Systems and Subsystems}

\subsection{Subject vs. object}

Tegmark devotes some introductory remarks in his article to an 
exposition of the philosophy in terms of which he means to account 
for cognition and, more specifically, its subjectivity.  This 
discussion rests on an extension of the usual decomposition of 
physical systems in terms of {\it object} and {\it environment} to 
include a third subsystem, the {\it subject}, consisting of the 
``degrees of freedom associated with the subjective perceptions of 
the observer,'' where the term `perception' is to include ``thoughts, 
emotions and any other attributes of the subjectively perceived state 
of the observer.''

By introducing the term `subject' in this context, Tegmark fails to 
remark an important conceptual distinction between subject and 
object.  In effect, he treats the subject as merely a special name 
for that object one studies when, for one reason or another, one 
wants to impute subjectivity to that collection of degrees of 
freedom, without explanation as to why these degrees of freedom in 
particular should have subjective implications or how they come to be 
associated with one another in a manner that does not depend on the 
arbitrary assignment of an observer.  It is not sufficient, in 
accounting for the fact that an observer subjectively perceives, to 
simply identify certain degrees of freedom as ``subjective.''  
Whereas the object is simply the name assigned to an arbitrarily 
delineated subsystem of the whole, the subject is not an arbitrary 
product of the way one happens to choose to analyze a system.  If it 
were, then it should be possible to associate a subject with any 
given subset of the available degrees of freedom.  Arbitrary 
collections of degrees of freedom are not, however, generally 
credited with subjective perception.  While it is a matter of the 
observer's choice what degrees of freedom to associate with an 
object, the subject must be determined as a matter of fact prior to 
any observer-orchestrated carving up of the problem.  The existence 
of an object of study is a relative fact, an artifact of analysis, 
whereas the existence of a subject is absolute, and its determination 
is a fact that is itself in need of explanation. 

Tegmark demonstrates his conception of the subject at work with a 
simple example involving two degrees of freedom, one an object, the 
other a subject.  If the object is in the state 
$\vert\!\uparrow\,\rangle$, the system involving both degrees of 
freedom evolves such that $U\vert 
{\raisebox{0pt}{..}\atop\raisebox{2pt}{-}}\!\uparrow\,\rangle
= \vert 
{\raisebox{0pt}{..}\atop\raisebox{2pt}{$\smile$}}\!\uparrow\,\rangle$ 
and likewise, if the object is in the state 
$\vert\!\downarrow\,\rangle$, the system evolves such that $U\vert 
{\raisebox{0pt}{..}\atop\raisebox{2pt}{-}} \!\downarrow\,\rangle
= \vert 
{\raisebox{0pt}{..}\atop\raisebox{2pt}{$\frown$}}\!\downarrow\,\rangle$.  
The joint subject/object density matrix, $\rho_{so}$, then evolves as:
\begin{eqnarray}
\rho_{so}&=&\frac{\mbox{\footnotesize 1}}{\mbox{\footnotesize 2}}\, 
U\Bigl( \vert {\raisebox{-3pt}{..}\atop\raisebox{7pt}{-}}
\rangle\langle {\raisebox{-3pt}{..}\atop\raisebox{7pt}{-}}
\vert \Bigr) \otimes \Bigl( 
\vert\!\uparrow\,\rangle\langle\,\uparrow\!\vert + 
\vert\!\downarrow\,\rangle\langle\,\downarrow\!\vert \Bigr) U^\dagger 
,\nonumber\\
&=&\frac{\mbox{\footnotesize 1}}{\mbox{\footnotesize 2}}\Bigl( \vert 
{\raisebox{-3pt}{..}\atop\raisebox{7pt}{$\smile$}}
\uparrow\,\rangle\langle 
{\raisebox{-3pt}{..}\atop\raisebox{7pt}{$\smile$}}
\uparrow\!\vert + \vert 
{\raisebox{-3pt}{..}\atop\raisebox{7pt}{$\frown$}}
\downarrow\,\rangle\langle 
{\raisebox{-3pt}{..}\atop\raisebox{7pt}{$\frown$}}
\downarrow\!\vert \Bigr) ,
\end{eqnarray}

Tegmark's interpretation of the final state as containing two 
definite but opposite subjective states correlated with the object 
state has no basis in the formalism.  Had he not illustrated the 
states in question with happy and sad faces, there would be nothing 
to necessitate, or even suggest, reading them as subjective 
perceptions.  In a less leading notation, 
\begin{equation}
\rho_{so}=\frac{\mbox{\footnotesize 1}}{\mbox{\footnotesize 2}}\Bigl( 
\vert\!\uparrow\,\uparrow\rangle\langle\uparrow\,\uparrow\!\vert + 
\vert\!\downarrow\,\downarrow\rangle\langle\downarrow\,\downarrow\!\vert 
\Bigr) ,
\label{eq:so}
\end{equation}
the same final state might be interpreted as a system in which there 
are simply two correlated object degrees of freedom.  The ambiguity 
of interpretation, and the fact that there need be no subjective 
implications whatsoever here, is indicative of the fact that the 
`subject' is simply a covert `object'.

In exploring the thesis that ``consciousness is synonymous with 
certain brain processes,'' Tegmark appears to mitigate this approach 
with the explicit recognition that consciousness is not arbitrarily 
allotted, but he gives us little help in understanding the 
particularity of these processes, or even why they should occur in 
brains.  His discussion in this context, of the mutual information, 
$I_{12}$, between the subject and its environment, might be construed 
as the tendering of a proposal to qualitatively distinguish, at least 
correlatively if not causally, the subject degrees of freedom that 
``{\it are} our perceptions.''  Yet his illustrations of a mutual 
information criterion involve only learning and information 
correlation of the kind seen above in the example of equation 
(\ref{eq:so}).  These concepts have no necessary connection to or 
implications for the subjective.  As Tegmark himself notes, his 
criterion would suggest that ``books and diskettes'' should have a 
subjective aspect to them, as would maps and road signs.

\subsection{The binding problem}

The conceptual confusion, in which the subject is treated as merely 
another kind of object, resurfaces in Tegmark's discussion of the 
binding problem.  The problem, as set out by James 
\cite{jam90,teg00}, refers to the fact that ``the only realities are 
the separate molecules, or at most cells.  Their aggregation into a 
`brain' is a fiction of popular speech.''\footnote{Of course, cells 
are as much physical aggregates as is the entire brain, and are just 
as aptly viewed as fictions.  James presumably concedes the existence 
of aggregates at a cellular level only to make contact with the 
fundamental units of biology and an audience of biologists.}  The 
brain is, in any physical description, merely an `object', in the 
sense discussed above, and is treated in this role as a unit merely 
by convention and not out of necessity.  While this has implications 
for most of the standard models in cognitive science, it is 
particularly damaging to an identity thesis, such as the one to which 
Tegmark explicitly subscribes.  If consciousness is synonymous with 
the brain, and the brain is merely a fiction of convenience, the 
inevitable conclusion is that consciousness is itself at best a 
fiction, a conclusion that we are all presumably in possession of 
enough personal evidence to reject.  To be sure, Tegmark speaks not 
of the brain {\it per se}, but of ``certain brain processes'' or of 
the ``subject degrees of freedom,'' taken collectively.  But 
processes and units comprised of degrees of freedom are as 
susceptible to James' complaint as is the physical substance of the 
brain -- none of these constitute, in a thoroughly classical 
understanding of cognition, more than a convenience.  Each disappears 
in a sufficiently fine analysis, replaced by a complex of purely 
local activities that, while perhaps more difficult to understand, 
are entirely adequate to the description of the physical goings-on.

What has been suggested by several commentators on the problem 
\cite{mar89,sta93,cai96} is that quantum coherence might account for 
holistic effects that thwart a purely local analysis, by introducing 
fundamentally non-local degrees of freedom.  While Tegmark apparently 
concedes the necessity for non-local binding in the determination of 
fundamental wholes, he finds non-local degrees of freedom aplenty in 
classical physics.  Crucially however, the degrees of freedom to 
which he points are merely artifacts of an approximate treatment.  
Oscillations of a guitar string, to borrow his example, can be 
treated as effectively non-local on timescales long compared to the 
timescale at which the forces that hold the string together propagate 
along the string.  Such accounts are sufficient to an `as if' account 
of the dynamics on the long timescales of an observer who lacks 
sufficient time resolution.  Nevertheless, we do not believe that 
such a treatment can give a {\it causal} description because it is 
not relativistic.\footnote{To put it another way, such a treatment is 
based on a `fictional' aggregate: the string.}  Tegmark fails to 
distinguish a fundamental non-locality from an effective non-locality 
in classical dynamics that arises only due to the presence of an 
insufficiently fine timescale, one that is associated with the 
shortcomings of an observer's knowledge of the system rather than 
with facts fundamental to an account of the ontology, aspects like 
causal propagation or the determination of a subject.  Thus, the 
conclusion that ``thoughts are presumably highly non-local excitation 
patterns in the neural network of our brain'' is a statement made 
from the perspective of an outside observer making convenient 
shorthand of the entirely local -- at the classical level of interest 
to Tegmark -- processes occurring in the extended space of the 
brain.  The introduction into classical description, of entities and 
levels of analysis that supersede the local level, is superfluous.

Tegmark attempts to make room for such levels of analysis by 
designating a `hyperclassical' class to distinguish the 
semiautonomous degrees of freedom  associated with the subject.  
These are identified with non-equilibrium, pumped and highly 
dissipative systems that do not conserve energy.  None of these 
criteria, however, constitute a qualitative distinction sufficient to 
remove hyperclassical degrees of freedom from the larger class of 
classical degrees in which they are contained, so that they appear 
equally subject to the criticisms above.  Moreover, Tegmark's 
characterization of these hyperclassical systems as those with 
$\tau_{\mathrm dec}\ll\tau_{\mathrm dyn}$ and a dissipation time 
$\tau_{\mathrm diss}\approx\tau_{\mathrm dyn}$, suggests that almost 
any open, classical system that is not adequately treated as 
independent can be hyper-classical.  Naturally, non-local degrees of 
freedom can be found, even in classical systems, at higher levels of 
complexity.  But these are not generally taken to be fundamental in a 
classical ontology, as they are inevitably tied to an observer's 
shortcomings in terms of resolution.

\section{Outlook}

As discussed in detail in this paper, none of the reasons that 
motivated a quantum approach to the problems peculiar to subjective 
states have been satisfactorily addressed within a classical 
framework in Tegmark's critique.  Neither do the mechanisms of 
decoherence discussed provide any clear evidence against the 
possibility of biologically instantiated quantum coherence of the 
sort envisioned in the Orch OR hypothesis.  Revisions to Tegmark's 
numerical estimates place the decoherence times of interest in a 
range that invalidates the assumptions from which the calculations 
proceeded and the low temperature limit suggests that the theoretical 
foundation is flawed.  When appropriately revised, both theoretically 
and numerically, decoherence times appear to be in line with 
appropriate dynamical times, an indication that there is cause for 
optimism that some of the fundamentally enigmatic features of the 
cognitive processes occurring in consciousness might yet be 
understood in the framework of a quantum theoretical solution.

\begin{acknowledgements}
S. H. and S. R. H. would like to thank Gerard Milburn for helpful 
discussions.  J. A. T. wishes to express his gratitude for research 
support from the  
Center for Consciousness Studies at the University of Arizona.  All  
authors are indebted to Roger Penrose for the formulation of the 
ideas under discussion and to Dave Cantrell for illustrations.
\end{acknowledgements}


\end{document}